\documentclass[10pt]{article}
\usepackage{psfig,epsf,conf-X_mod}
\begin{document} 
\small
\heading{Cosmology with SZ and X-ray cluster surveys\footnote{ Contribution to ``Large Scale Structure in the X-ray Universe'' Workshop, Santorini, Greece, September 1999, eds. M. Plionis and I. Georgantopoulos (Editions Frontieres)}}
\par\medskip\noindent
\author{R\"udiger Kneissl}
\address{Cavendish Astrophysics, Department of Physics, 
University of Cambridge,\\
Madingley Road, Cambridge CB3 0HE, UK}
\begin{abstract}
Hydrodynamical simulations are used in combination with the 
Press-Schechter expression to simulate Sunyaev-Zel'dovich 
(SZ) galaxy cluster sky maps. These are used to gauge 
the ability of future SZ observations to provide 
information about the cosmological parameters $H_0$, $\Omega_{\mathrm 
M}$, and the gas fraction $f_g$ in clusters. 
This work concentrates on prospects for AMI, the Arcminute 
MicroKelvin Imager, a new type of compact interferometric array 
currently proposed in Cambridge. The expectations are contrasted 
with those for X-ray missions, such as XMM, and the benefits of 
combining SZ and X-ray data are highlighted.
\end{abstract}
\section{AMI and SZ cluster skys}
The Sunyaev-Zel'dovich (1972, SZ) effect, unlike optical and X-ray 
cluster surveys, is not affected by redshift because it measures the 
integrated line of sight intracluster gas pressure via its Compton scattering 
of cosmic microwave background (CMB) photons. During the last 10 years, 
interferometric techniques have been developed, which are providing 
firm detections of known clusters (eg. Jones et al. 1993, 
Carlstrom, Joy and Grego 1996). The technology and expertise is now 
available to survey the sky to discover clusters. 

The construction of a CMB telescope for the observation of the 
CMB on angular scales of one to several arcminutes and with a 
sensitivity of a few micro-Kelvin, comparable to the Planck 
Surveyor, has been proposed in Cambridge. The Arcminute 
MicroKelvin Imager (AMI) will consist of a small compact array 
of 4 meter dish antennas combined with the existing Ryle 
telescope antennas in an extended array, with a new receiving 
system and novel correlator. This design achieves 
optimal sensitivity to the cluster SZ effect and a separation 
with other components such as the primary CMB and radio sources. 
Although the instrument is dedicated to the study of clusters, 
it will generally probe the structure of the CMB on sub-Planck 
and super-ALMA scales. Therefore it is also sensitive to other phenomena 
such as, inhomogeneous ionisation, density - velocity correlations 
(Vishniac and Ostriker effect), filaments and topological defects, 
if they exist, which are all of immense interest as well. In the 
following however we demonstrate the ability of AMI to discover clusters. 

To produce simulated SZ cluster sky maps, 
the Press-Schechter expression (1974) is used to create a list of
cluster masses and redshifts having an abundance consistent with the
local cluster temperature function (Eke et al. 1996). These clusters
are placed at random angular positions within a 5$^\circ$ $\times$
5$^\circ$ sky map with 40 arcsec pixels. To model the cluster SZ
signal, template maps have been created from the hydrodynamical
simulations of Eke, Navarro and Frenk (1998), and these are pasted,
suitably scaled, onto the cluster positions. This procedure is performed 
for two cases, a low present density ($\Omega = 0.3$) and a high density 
($\Omega = 1$) universe, both with a Hubble constant of 
70 km s$^{-1}$ Mpc$^{-1}$. The gas fraction is fixed at 10~\% 
in both cases rather than to the primordial nucleosynthesis value, 
which would have introduced an $\Omega$ dependence, enhancing the 
differences. Our value for the gas fraction is at the 
low end of the value estimated from X-ray clusters (Ettori and Fabian (1999) 
and Mohr, Mathiesen and Evrard (1999) find it to be 0.1-0.25 at the 95 \% CL), 
and the model SZ number counts would increase with a less conservative 
choice. We also produced corresponding X-ray maps in 0.5-2 and 2-10 
keV bands, and have checked that they are complete to an X-ray flux limit of 
$1 \times 10^{-15}$ erg cm$^{-2}$ s$^{-1}$ [0.5-2 keV]. 
\section{An SZ survey and X-ray / optical follow-up}
The sensitivity of future instruments, we use AMI specific numbers, 
is high enough to allow a survey of the sky for SZ clusters. 
To make our simulation of the observation process 
computationally feasible we simplified the observation by running a 
compensated beam with carefully chosen shape and amplitude over the 
cluster map. This procedure had been gauged with detailed simulations 
of the interferometer response including the specific noise properties 
of a field observation and through the recovery process. The resulting 
number counts are shown in figure 1. 
\begin{figure}[htb]
\mbox{\epsfxsize=8cm \epsffile{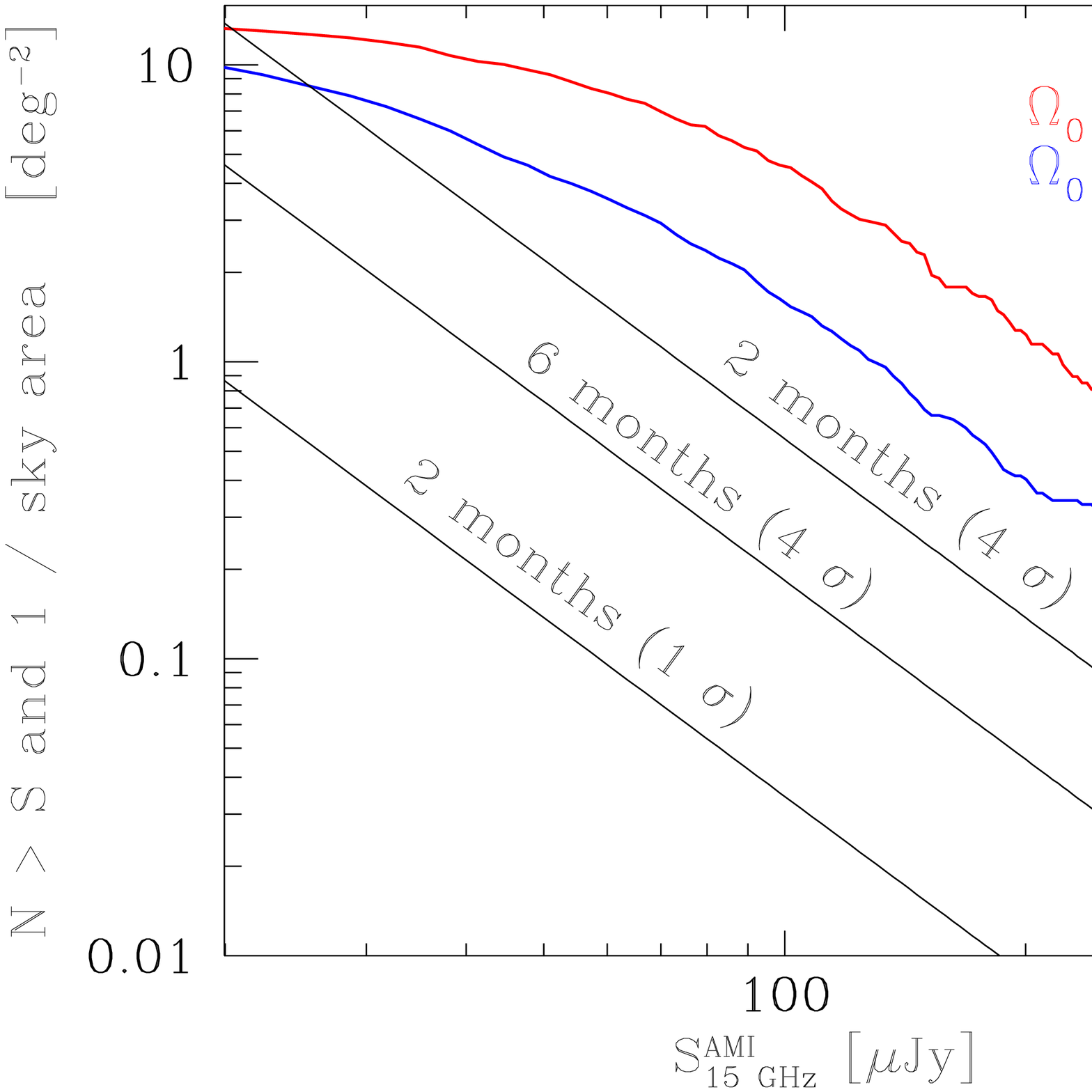}}
\caption[]{SZ number counts and AMI sensitivity}
\end{figure}
There is a turn-over at the confusion limit for a 4.5 arcmin beam. 
Also shown in figure 1 are sensitivity lines each corresponding to a 
fixed observation time and displayed as the inverse of the survey area 
against the flux limit. The ratio between the number counts and the 
sensitivity lines is the number of clusters detected in the respective time. 
The maximum detection number lies at a limiting observed flux of about 
100 $\mu$Jy or a corresponding survey area of 5 deg$^2$ and about 10 
clusters for the high density case and tens of clusters for low density are 
expected for a 6 months observation. Since the optimum is shallow 
it will also be possible to adopt different surveying strategies at a 
low cost of inefficiency. This will probe the slope of the number counts 
which is a function of $\Omega$. 
The parameters of the simulations, which have the largest effects 
on the number counts are listed with present uncertainties 
in table 1. Their effect is considerable, but substantial improvements 
can be expected in the future, in particular from the new X-ray 
missions. 

Due to the $(1+z)^4$ dimming of bolometric X-ray flux 
compared to the Compton scattering process the X-ray and SZ flux limited 
cluster samples have very different redshift distributions, 
even for an X-ray limit of $5 \times 10^{-15}$ erg cm$^{-2}$ s$^{-1}$ 
[0.5-2 keV] in the case of the currently favoured low density model. 
With scaling laws we find that the 
ratio between SZ to X-ray flux scales with $(1+z)^{5/2}$ and 
with only a weak dependence on $\Omega$ and the cluster temperature. 
When observing the same field in microwaves and X-rays this "photometric" 
redshift effect will immediately allow a crude measure of the 
redshift distribution (see figure 2). 
\begin{figure}[htb]
\mbox{\epsfxsize=9cm \epsffile{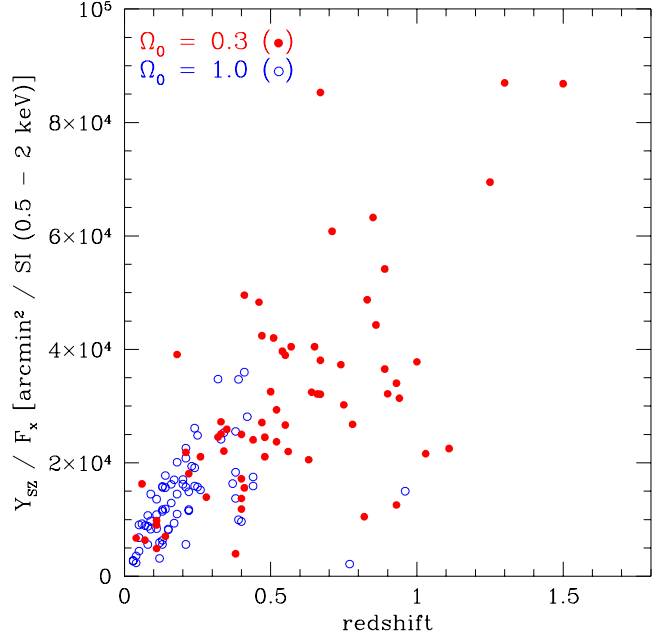}}
\caption[]{SZ - X-ray flux ratio against redshift for an SZ flux 
limited sample}
\end{figure}
A limit on the redshift will be 
obtained for the SZ clusters undetected in X-rays. 
A major uncertainty in determining $\Omega$ from this distribution comes 
from $f_g$. However when optical redshifts are obtained 
we find that for the flux limit reached with AMI the median redshift of 
only 20 clusters will allow to distinguish between the two cases with more 
than 99~\% confidence. Going back to the X-ray and SZ data with an estimate 
of $\Omega$, the gas physics for example $f_g(z)$ can be studied. 
\begin{center}
\begin{tabular}{ccc}
\multicolumn{3}{l}{Table 1: Model parameters affecting SZ number counts} \\
\hline
parameter & change in percent & fractional change in $N(>Y)$ \\
\hline
$h$ & 20 \% & 1.3 \\
$f_g$ & 30 \% & 1.5 \\
$\sigma_8$ & 7 \% (1$\sigma$)& 1.5 \\
 & 14 \% (2$\sigma$) & 3.2 \\
\hline
\multicolumn{3}{l}{Compare this to $N(\Omega=0.3) / N(\Omega=1) 
\approx 3.5$}
\end{tabular}
\end{center}
\section{Conclusions}
Instruments to survey the sky for SZ clusters can be build. 
The expected number of clusters is greater than 10 for 
a half year observation and depends strongly on the matter density. 
The SZ data in combination with X-ray and optical data can be used 
to constrain $\Omega, f_g$, $\sigma_8$, and from distance measurements, 
$H_0$ and most likely $\Omega_\Lambda$, if enough suitable clusters are 
present at high redshift. 
SZ cluster surveys will be particularly useful as pathfinders 
for future X-ray missions such as Constellation-X and XEUS, 
which will have small fields of view. With the cluster SZ effect 
a sample of massive clusters at high redshift can be provided as 
targets for a detailed study of the plasma physics in X-rays, 
interesting for the understanding of cluster formation and cosmology. 

\acknowledgements{I am grateful to my collaborators 
Mike Jones and {\mbox Vincent} Eke, and acknowledge financial support 
from an EU Marie Curie Fellowship.}

\vfill
\end{document}